\documentclass[prl, twocolumn, superscriptaddress,floatfix]{revtex4}%

\usepackage{amsfonts}
\usepackage{amsmath}
\usepackage{amssymb}
\usepackage{graphicx}
\usepackage{epstopdf}
\usepackage{textcomp}
\usepackage[normalem]{ulem}
\usepackage{color}
\usepackage{gensymb}

\newcommand{\SnxInxTe}{Sn$_{1-x}$In$_x$Te}

\begin{document}

\preprint{}

\title{Superconductivity, pairing symmetry, and disorder in the doped topological insulator Sn$_{1-x}$In$_x$Te for x $\geq$ 0.10}

\author{M.~P.~Smylie}
\affiliation{Materials Science Division, Argonne National Laboratory, 9700 S. Cass Ave., Argonne, Illinois 60439}
\affiliation{Department of Physics, University of Notre Dame, Notre Dame, Indiana 46556}
\author{H.~Claus}
\affiliation{Materials Science Division, Argonne National Laboratory, 9700 S. Cass Ave., Argonne, Illinois 60439}
\author{W.-K.~Kwok}
\affiliation{Materials Science Division, Argonne National Laboratory, 9700 S. Cass Ave., Argonne, Illinois 60439}
\author{E. R. Louden}
\affiliation{Department of Physics, University of Notre Dame, Notre Dame, Indiana 46556}
\author{M.~R.~Eskildsen}
\affiliation{Department of Physics, University of Notre Dame, Notre Dame, Indiana 46556}
\author{A. S. Sefat}
\affiliation{Materials Science and Technology Division, Oak Ridge National Laboratory, Oak Ridge, Tennessee 37831}
\author{R. D. Zhong}
\affiliation{Condensed Matter Physics and Materials Science Department, Brookhaven National Laboratory, Upton, New York 11973}
\affiliation{Materials Science and Engineering Department, Stony Brook University, Stony Brook, New York 11749}
\author{J. Schneeloch}
\affiliation{Condensed Matter Physics and Materials Science Department, Brookhaven National Laboratory, Upton, New York 11973}
\affiliation{Department of Physics and Astronomy, Stony Brook University, Stony Brook, New York 11973}
\author{G. D. Gu}
\affiliation{Condensed Matter Physics and Materials Science Department, Brookhaven National Laboratory, Upton, New York 11973}
\author{E.~Bokari}
\affiliation{Department of Physics, Western Michigan University, Kalamazoo, Michigan 49008}
\author{P.~M.~Niraula}
\affiliation{Department of Physics, Western Michigan University, Kalamazoo, Michigan 49008}
\author{A.~Kayani}
\affiliation{Department of Physics, Western Michigan University, Kalamazoo, Michigan 49008}
\author{C. D. Dewhurst}
\affiliation{Institut Laue-Langevin, 6 Rue Jules Horowitz, F-38042 Grenoble, France}
\author{A.~Snezhko}
\affiliation{Materials Science Division, Argonne National Laboratory, 9700 S. Cass Ave., Argonne, Illinois 60439}
\author{U.~Welp}
\affiliation{Materials Science Division, Argonne National Laboratory, 9700 S. Cass Ave., Argonne, Illinois 60439}

\begin{abstract}
The temperature dependence of the London penetration depth $\Delta\lambda(T)$ in the superconducting doped topological crystalline insulator Sn$_{1-x}$In$_x$Te was measured down to 450 mK for two different doping levels, x $\approx$ 0.45 (optimally doped) and x $\approx$ 0.10 (underdoped), bookending the range of cubic phase in the compound.
The results indicate no deviation from fully gapped BCS-like behavior, eliminating several candidate unconventional gap structures.
Critical field values below 1 K and other superconducting parameters are also presented.
The introduction of disorder by repeated particle irradiation with 5 MeV protons does not enhance $T_c$, indicating that ferroelectric interactions do not compete with superconductivity.
\end{abstract}

\date{\today}

\maketitle

\begin{center}
\textbf{I. INTRODUCTION}
\end{center}

Recently, there has been significant attention given to topological states in solids, particularly towards topological insulators (TI) \cite{Hasan-Kane-RevModPhys-Review-of-TI,Ando-JPhysJpn-TI-review} and topological superconductors (TSC) \cite{Qi-RevModPhys-Review-of-TSC-theory,Sasaki-Mizushima-PhysicaC-STI-review}, because of the properties of their novel quantum states.
A topological insulator is a material that is insulating in the bulk, but has gapless surface states that conduct; these states are protected by time-reversal symmetry in the material.
In topological crystalline insulators (TCIs) \cite{Fu-PRL-TCI}, the gapless surface state is instead protected by the mirror symmetry of the crystal.
Following confirmation of Bi$_2$Se$_3$, Bi$_2$Te$_3$, and Sb$_2$Te$_3$ as topological insulators, a few materials have been identified as topological crystalline insulators \cite{Ando-Fu-AnnuRevCMP-TCI-and-TSC} including Pb$_{1-x}$Sn$_x$Se, Pb$_{1-x}$Sn$_x$Te, and SnTe \cite{Hsieh-Fu-NatComm-SnTe-TCI,Sato-Ando-PRL-Fermiology-SnInTe}.
Topological superconductors support gapless surface quasiparticle states that can host Majorana fermions, whose non-Abelian statistics may form the basis for new approaches to fault-tolerant quantum computing \cite{Wilczek-QC,Beenakker-AnnRevCMatt-Majoranas-in-TSCs,Albrecht-Nature-Zero-modes-majoranas,Mourik-Science-Majoranas-in-Nanowires}.
Two routes are currently being pursued \cite{Qi-RevModPhys-Review-of-TSC-theory,Sasaki-Mizushima-PhysicaC-STI-review,Ando-Fu-AnnuRevCMP-TCI-and-TSC,SatoAndo-RepProgPhys-TSC-review} to create a topological superconductor: proximity induced at the interface between strong spin-orbit coupling semiconductors and conventional superconductors, or by chemical doping of bulk TI and TCI materials.
Among the latter, the first materials suggested to be bulk topological superconductors were obtained by doping Bi$_2$Se$_3$: Cu$_x$Bi$_2$Se$_3$ \cite{Hor-Cava-PRL-CBS-discovery,Das-PRB-triplet-CBS,KrienerAndo-PRL-CBS-specific-heat,Kriener-Ando-PRB-CBS-synthesis,Schneeloch-PRB-CBS-synthesis} with $T_c \sim$ 3.5 K, Nb$_x$Bi$_2$Se$_3$ with $T_c \sim$ 3.4 K and Sr$_x$Bi$_2$Se$_3$ \cite{Hor-NBS-arxiv,Smylie-PRB-NBS-TDO,Liu-JACS-SBS-discovery,Shruti-PRB-SBS-characterization,Wang-ChemMat-SC-in-TlBT} with $T_c \sim$ 3.0 K.
More recently, surface Andreev bound states in In-doped SnTe crystals have been observed \cite{Sasaki-Fu-Ando-PRL-PCS-in-SnInTe} via point-contact spectroscopy; the presence of such zero-bias conductivity peaks are generally interpreted as sign of unconventional superconductivity \cite{Kashiwaya-ZBCP-is-unconventional-SC}.
Thermal conductivity measurements \cite{He-PRB-SnInTe-thermal-conductivity} on a Sn$_{0.6}$In$_{0.4}$Te crystal suggest a full gap, and Knight shift measurements \cite{Maeda-Ando-PRB-SnInTe-NMR-says-not-triplet} on a polycrystalline sample with $\sim$4\% doping may indicate a spin-singlet state.
In systems with time reversal and inversion symmetry, odd-parity pairing is a requirement for topological superconductivity.
Thus, determining the superconducting gap structure is important to establishing the possibility of topological superconductivity, as not all theoretically allowed \cite{Sasaki-Fu-Ando-PRL-PCS-in-SnInTe} gap structures are unconventional, odd-parity states.

The phase diagram of \SnxInxTe~is known to contain several phases \cite{Novak-Ando-PRB-SnInTe-phase-diagram}.
The parent compound SnTe undergoes a ferroelectric transition at up to 100 K; this transition temperature decreases to zero with increasing hole concentration \cite{Erickson-Fisher-PRB-SnTe-SC-2009}.
The ferroelectric transition is accompanied by a structural phase change from cubic to rhombohedral.
At sub-Kelvin temperatures, the parent material becomes superconducting \cite{Hein-PR1966-SC-in-SnTe,Mathur-JPhysChemSol1973-SC-in-SnTe}.
It was discovered that In-doping on the Sn site increases the superconducting transition temperature by an order of magnitude, a surprising result considering its low carrier density of $\sim 10^{21}$ cm$^{-3}$.
More recent efforts \cite{Zhong-PRB2013-Optimize-Tc-in-SnInTe,Balakrishnan-PRB2013-SC-properties-of-SnInTe}, spurred by the growing interest in topological materials, have raised the transition temperature in \SnxInxTe~to 4.5 K with better synthesis techniques.
The low-temperature phase diagram is separated into two crystal structures: for x $<$ 0.04, the structure is rhombohedral, and for x $>$ 0.04, the structure is face centered cubic.
For a narrow range of doping (0.02 $<$ x $<$ 0.04), the compound \SnxInxTe~is both ferroelectric and superconducting, both of which are thought to be bulk in nature.
In this range, $T_c$ is below 2 K and is not a function of x \cite{Novak-Ando-PRB-SnInTe-phase-diagram}.
Above this range, up to the solubility limit of x $\sim$ 0.45, $T_c$ increases linearly with x to a maximum of $\sim$ 4.5 K.
Recent reports suggest \cite{Novak-Ando-PRB-SnInTe-phase-diagram,HaldolaarachchigeCava-PRB2016-SnInTe-phase-diagram} that the pairing mechanism may be different for low and high doping levels, and that disorder scattering may have a strong effect on the transition temperature.
In as-grown crystals shown to have equal carrier concentrations \cite{Novak-Ando-PRB-SnInTe-phase-diagram}, crystals with higher normal-state resistivity systematically have higher $T_c$'s.
This may be due to either disorder favoring even pairing channels over odd \cite{Kozii-Fu-PRL2015-competing-odd-and-even-parity,Wu-Martin-PRB2017-odd-and-even-parity-competition-nematic} or by favoring superconducting over ferroelectric interactions.

In this work, we report on magnetization measurements and low-temperature measurements of the London penetration depth $\lambda$.
The temperature dependence of $\lambda$ indicates a full superconducting gap.
Increased electron scattering induced by particle irradiation does not enhance $T_c$ in the cubic phase of \SnxInxTe~implying that for higher doping levels, the competition between ferroelectric, odd-parity, and even parity is weak if extant, as odd-parity pairing is conventionally thought to be very sensitive to nonmagnetic disorder \cite{BalianWerthamer-PR1963-pwave}.

\begin{center}
\textbf{II. EXPERIMENTAL METHODS}
\end{center}

Crystals of Sn$_{0.9}$In$_{0.1}$Te and Sn$_{0.55}$In$_{0.45}$Te were grown by the modified Bridgman method, following the work of Tanaka \cite{Tanaka-NatPhys-SnTe-is-a-TCI}.
This range of x was chosen to cover the range of the cubic superconducting phase while remaining clearly above the cubic-rhombohedral structural transition.
X-ray diffraction and EDS measurements were used to verify the crystal structure and stoichiometry.

Magnetometry measurements were performed both with a Quantum Design MPMS dc SQUID magnetometer with a superconducting magnet down to 1.8 K, and a custom-built SQUID magnetometer with a conventional magnet down to 1.2 K.
The tunnel diode oscillator (TDO) technique \cite{Prozorov-SST2006-penetration-depth-review,Prozorov-RepProgPhys2011-penetration-depth-FeSC} was used to measure the temperature dependence of the London penetration depth $\Delta\lambda(T) = \lambda(T) - \lambda_0$ with $\lambda_0$ the zero-temperature value in various applied magnetic fields down to 400 mK in a $^3$He cryostat with a custom \cite{Smylie-PRB-NBS-TDO,Shen-Smylie-PRB2015-BaK122,Smylie-PRB2016-BaFeAsP-irradiation} resonator operating at $\sim$14.5 MHz.
To image the vortex lattice in the superconducting state and to obtain an independent estimate of $\lambda_0$, complementary small-angle neutron scattering (SANS) measurements were performed at 50 mK on the D33 beam line at the Institut Laue-Langevin in Grenoble, France \cite{Eskildsen-SnInTe-ILL-DOI-Citation}.
To examine the role of disorder, repeated irradiation with 5-MeV protons was performed at the tandem Van de Graaf accelerator at Western Michigan University.
Irradiation with MeV-energy protons creates a distribution of defects, ranging from Frenkel pairs of point defects to collision cascades and clusters \cite{Kirk-Micron1999-YBCO-irradiation-defects,LeiFang-PRB2011-BaFeAsP-irradiation,Civale-PRL1990-irradiated-YBCO}, all of which enhance electron scattering.
During irradiation the samples were cooled to approximately -10 $\degree$C to prevent local heating of the sample.

\begin{center}
\textbf{III. RESULTS AND DISCUSSION}
\end{center}

\begin{figure}
	\includegraphics[width=1\columnwidth]{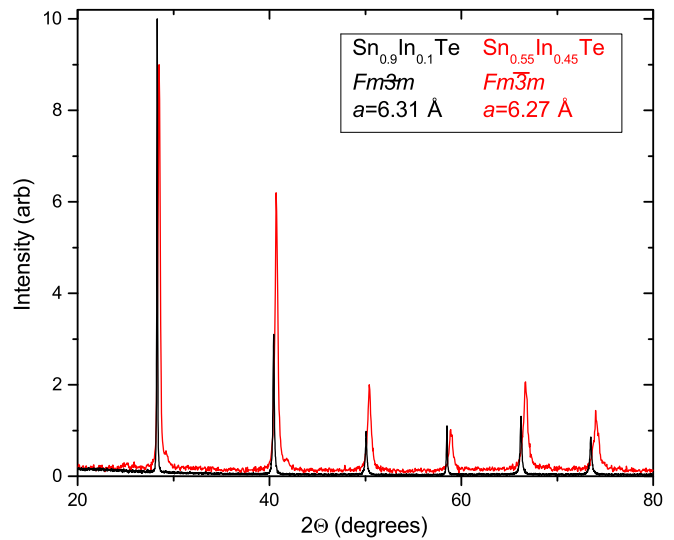}
	\caption{
		Diffraction data verifying single-phase FCC structure in underdoped and optimally doped \SnxInxTe.
		In both materials, the lattice constants are consistent with the doping levels as measured through EDS analysis.
	}
	\label{figXRD}
\end{figure}

X-ray diffraction measurements verifying the crystal structure and purity are shown in Fig. \ref{figXRD}.
At both doping levels, the material is single phase with rocksalt structure (space group $Fm\bar{3}m$), with lattice parameters $a$ = 6.31 \AA~for x $\approx$ 0.1, and $a$ = 6.27 \AA~for x $\approx$ 0.45.
Through EDS analysis the composition was determined, yielding values close to the nominal stoichiometry.

\begin{figure}
	\includegraphics[width=1\columnwidth]{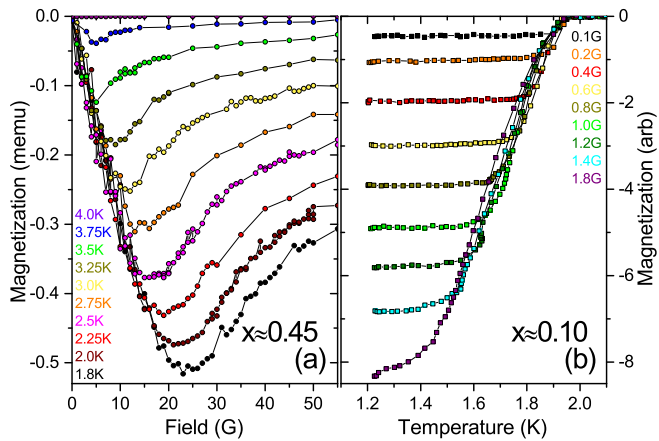}
	\caption{
		(a) Magnetization versus applied magnetic field sweeps on optimally doped \SnxInxTe~ at various temperatures from 1.8 K through 4.0 K, measured in a conventional MPMS SQUID.
		(b) Magnetization versus temperature sweeps on 10\% doped \SnxInxTe at various fields from 0.1 G to 1.8 G, measured in a custom dc SQUID.
		Isothermal magnetization versus field curves are extracted from this data.
		In both datasets, $H_p$ is determined as the field for which the magnetization deviates away from being linear in H.
	}
	\label{figHc1-sweeps}
\end{figure}

Estimates of $\lambda_0$ can be obtained from measurements of the lower critical field $H_{c1}$ and upper critical field $H_{c2}$.
Values of $H_{c1}$ for both doping levels were deduced from low-temperature magnetization measurements shown in Fig. \ref{figHc1-sweeps}.
For the optimally doped material, magnetization measurements versus applied field [Fig. \ref{figHc1-sweeps}(a)] were used; for the x $\approx$ 0.1 material, magnetization versus temperature measurements at multiple fixed fields in the range of 0.1-1.8 G [Fig. \ref{figHc1-sweeps}(b)] were performed, and magnetization versus applied field could be extracted from isothermal data.
In both cases, the penetration field $H_p$ \cite{Shen-Smylie-PRB2015-BaK122,Brandt2013-JETP-Brandt-Hp} was taken as the field for which the magnetization deviates away from being linear in H.
Using the Brandt formulation \cite{Brandt2013-JETP-Brandt-Hp}, we calculate the corrections due to edge and/or surface barriers to vortex penetration yielding estimates of $H_{c1}$ as shown in Fig. \ref{figHc1Hc2}.
For a platelike superconductor, $H_p / H_{c1} = \textnormal{tanh}\left(\sqrt{\alpha t/w}\right)$, where $t$ and $w$ are the thickness and width, and $\alpha$ = 0.67 for a disc-shaped sample.
Upper and lower critical field data for both doping levels are shown in Fig. \ref{figHc1Hc2}.
With a conventional parabolic temperature dependence $H_{c1} = H_{c1}(0)\left(1-(T/T_c)^2\right)$ we extrapolate $H_{c1}$ = 7.96 G and 32.0 G as the zero-temperature values for x $\approx$ 0.1 and x $\approx$ 0.45, respectively.

\begin{figure}
	\includegraphics[width=1\columnwidth]{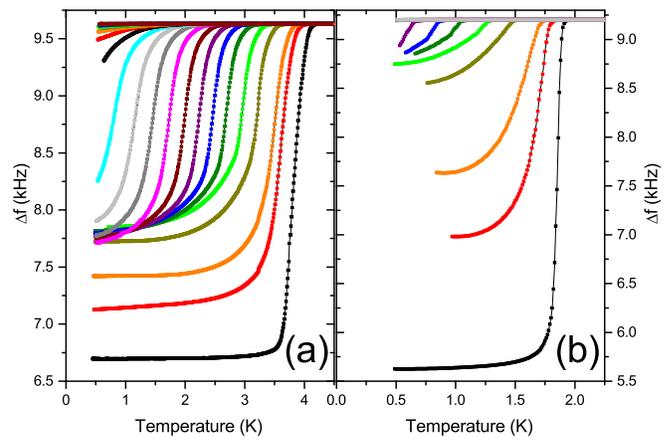}
	\caption{
		TDO measurements showing suppression of superconductivity in applied magnetic fields up to 2.1 T for near-optimally doped (a) and underdoped (b) \SnxInxTe.
	}
	\label{figTDOcomposite}
\end{figure}

\begin{figure}
	\includegraphics[width=1\columnwidth]{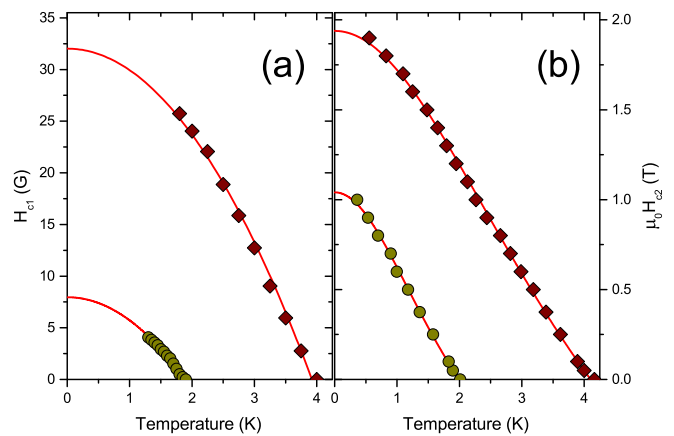}
	\caption{
		Critical field $H_{c1}$ (a) and $H_{c2}$ (b) values for underdoped (yellow circles) and near-optimal (red diamonds) \SnxInxTe.
		Extrapolated zero-temperature values for $H_{c1}$ are 7.92 G and 32.0 G, and for $H_{c2}$ 1.04 T and 1.94 T, for underdoped and near-optimal, respectively.
	}
	\label{figHc1Hc2}
\end{figure}

The TDO frequency shift is proportional to the magnetic susceptibility \cite{Prozorov-SST2006-penetration-depth-review,Prozorov-RepProgPhys2011-penetration-depth-FeSC} of the sample, allowing for the detection of the superconducting transition as shown in Fig. \ref{figTDOcomposite} for field values up to 2 T for small crystals of both doping levels.
No secondary superconducting transitions were observable in either sample up to 20 K.
Defining the onset $T_c$ to be at the deviation in slope of the TDO frequency shift from the essentially temperature independent value at temperatures above $T_{c0}$ yields the H$_{c2}$(T) data shown in Fig. \ref{figHc1Hc2}.
The phenomenological relation $H_{c2}$(T) = $H_{c2}(0)\left(\frac{1-t^2}{1+t^2}\right)$, shown in red in Fig. \ref{figHc1Hc2}, describes the data well, as has been observed for other superconducting doped topological insulators \cite{Nikitin-PRB-SBS-basal-anisotropy-pressure}.
This yields a zero-temperature limit of the upper critical field $H_{c2}$ of approximately 1.04 T for the underdoped sample, and for the near-optimally doped sample, $H_{c2}(0) \approx$ 1.94 T.
Both values are well below the BCS Pauli paramagnetic limit of B$_{c2}^{Pauli} = 1.83 T_c$.
From our values of $H_{c2}$, we calculate the coherence length $\xi_0$ for both doping levels using the Ginzburg-Landau relation $\mu_0 H_{c2}(0) = \Phi_0/2\pi\xi^2(0)$, resulting in $\xi_0$ = 17.8 nm for x $\approx$ 0.1 and $\xi_0$ = 13.0 nm for x $\approx$ 0.45.
With the extrapolated zero-temperature $H_{c2}$ values and using the Ginzburg-Landau formula $H_{c1} = \Phi_0/(4\pi\lambda^2)(ln[\lambda/\xi] + 0.5)$, we determine estimates for the zero-temperature value of $\lambda$ to be 900 nm for x $\approx$ 0.1 and 425 nm for x $\approx$ 0.45; such large values are consistent with values from NMR \cite{Maeda-Ando-PRB-SnInTe-NMR-says-not-triplet} ($\sim$1200 nm, x = 0.04) and $\mu$SR \cite{Saghir-Balakrishnan-PRB2014-SnInTe-muSR} (542 nm, x = 0.4).

SANS measurements were performed on oriented crystals of Sn$_{0.9}$In$_{0.1}$Te.
Data was collected at 50 mK for applied magnetic fields ranging from 0.1 to 0.3 T directed along various high symmetry directions, but no vortex lattice could be detected.
From the background intensity, a lower limit of the London penetration depth $\lambda_0$ may be extracted from the neutron reflectivity R:
\begin{equation}
\centering
	R = \frac{2\pi\gamma_n^2}{16\phi_0^2} \frac{t\gamma^2}{q} \frac{B^2}{(1+\lambda^2q^2)^2} ~\textnormal{exp} \left(-2c\xi^2q^2\right)
\end{equation}
where $\gamma_n$ is neutron gyromagnetic ratio, $t$ the sample thickness, $B$ the applied magnetic field, $\phi_0$ = 2067 T nm$^2$ the flux quantum, $q$ the scattering vector, and $\xi$ the coherence length, with $c$ a constant typically taken as 0.5 \cite{Eskildsen-FrontPhys-SANS-review}.
Our SANS results put a lower limit of 550 nm on $\lambda_0$, consistent with our direct estimate of $\lambda_0$ via lower and upper critical fields.

Low temperature penetration depth measurements were carried out via the TDO technique in the temperature range from 0.4 to 40 K.
In the TDO technique, the frequency shift $\delta f$ of the resonator is proportional to the change of the penetration depth \cite{Prozorov-RepProgPhys2011-penetration-depth-FeSC}:
\begin{equation}
	\delta f(T) = G\Delta\lambda(T)
\end{equation}
where the geometrical factor $G$ depends on the sample shape and volume as well as the geometry of the resonator coil.
The magnetic field of the resonator coil is $\sim$20 mOe, assuring that the sample remains fully in the Meissner state.

The low-temperature variation of the London penetration depth $\Delta\lambda (T) = \lambda(T) - \lambda_0$ can provide information on the superconducting gap structure \cite{Prozorov-SST2006-penetration-depth-review}.
In the low temperature limit, conventional BCS theory for an isotropic $s$-wave superconductor yields an exponential variation of $\Delta\lambda(T)$:
\begin{equation}
	\frac{\Delta\lambda(T)}{\lambda_0} \approx \sqrt{\frac{\pi\Delta_0}{2T}}~ \textnormal{exp}\left(\frac{-\Delta_0}{T}\right)
	\label{BCS-gap-lambda}
\end{equation}
with $\lambda_0$ and $\Delta_0$ the zero-temperature values of the penetration depth and energy gap.
In contrast, in nodal superconductors the enhanced thermal excitation of quasiparticles near the gap nodes results in a power law variation, $\Delta\lambda \sim T^n$ \cite{Smylie-PRB-NBS-TDO,Prozorov-SST2006-penetration-depth-review,Gross-Hirschfeld-ZfurPhysB1986-lambda-in-triplets} where the exponent $n$ depends on the nature of the nodes and the degree of electron scattering.

\begin{figure}
	\includegraphics[width=1\columnwidth]{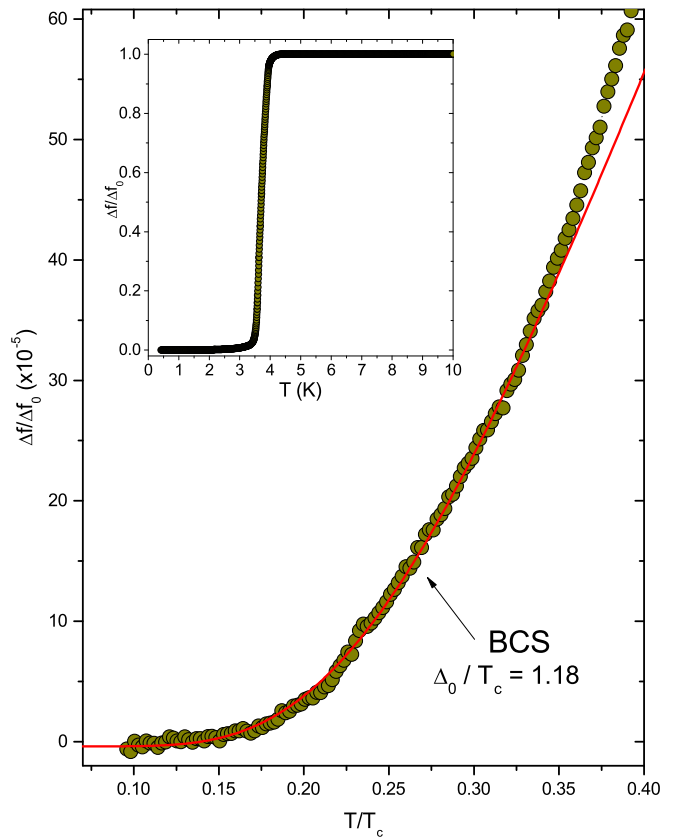}
	\caption{
		Normalized low temperature frequency shift $\Delta f(T)$ for Sn$_{0.55}$In$_{0.45}$Te.
		The BCS-like fit (red) well describes the data.
		The inset shows the full, sharp transition, with no evidence of other low-temperature phases.
	}
	\label{fig45TDO}
\end{figure}

The evolution of the low temperature TDO response of a single crystal of Sn$_{0.55}$In$_{0.45}$Te is shown in Fig. \ref{fig45TDO}.
The inset shows the full transition, which is very sharp, indicating a high quality material.
The behavior of the optimally doped material can be well described by an exponential dependence with a BCS-like gap value (red line) below $T_c$/3, indicating that the material is a fully-gapped superconductor, in agreement with thermal conductivity and muon-spin spectroscopy measurements \cite{He-PRB-SnInTe-thermal-conductivity,Saghir-Balakrishnan-PRB2014-SnInTe-muSR}.
Our data extend a recent report \cite{Maurya-Shruti-EPL2014-SnInTe-useless-TDO-data} to low temperatures where Eq. \ref{BCS-gap-lambda} is actually applicable.
The low gap ratio of $\Delta_0/T_c = 1.18$ is not consistent with standard BCS $s$-wave theory which predicts $\Delta_0/T_c = 1.76$, but is consistent with a weakly anisotropic single gap \cite{Manzano-Carrington-PRL2002-MgB2-anisotropic-swave,CarringtonGiannetta-PRL2007-NbSe2-penetration-depth,MarsiglioCarbotte-PhysofSC} as the temperature dependence of $\lambda$ probes quasiparticle excitations at the lowest activation energy.

\begin{figure}
	\includegraphics[width=1\columnwidth]{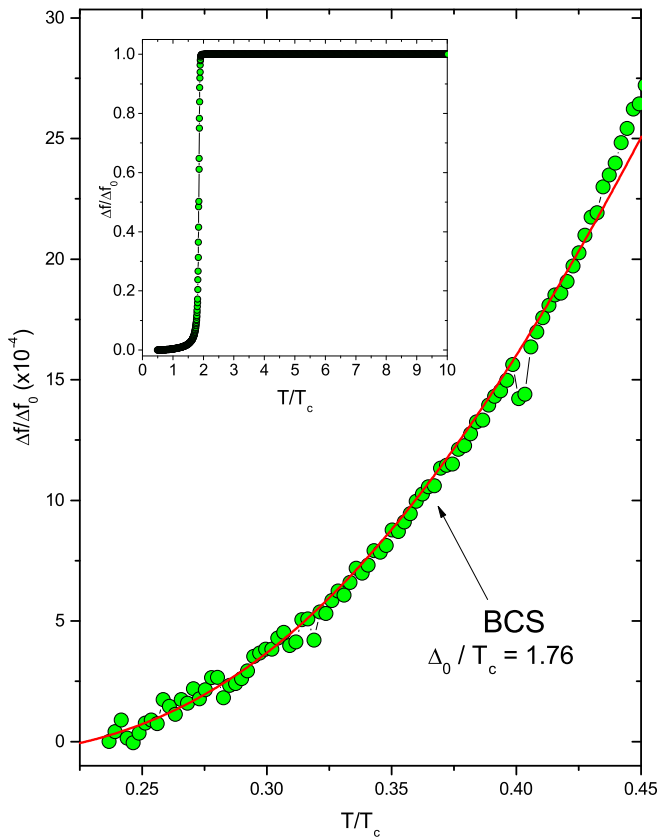}
	\caption{
		Normalized low temperature frequency shift $\Delta f(T)$ for Sn$_{0.9}$In$_{0.1}$Te.
		The BCS-like fit (red) well describes the data.
		The inset shows the full, sharp transition, with no evidence of other low-temperature phases.
	}
	\label{fig10TDO}
\end{figure}

The x $\approx$ 0.1 doping level is slightly above the value separating the ferroelectric rhombohedral phase and the cubic phase.
The low temperature TDO response for a single crystal of Sn$_{0.9}$In$_{0.1}$Te is shown in Fig. \ref{fig10TDO}.
The inset shows the full transition, which is very sharp.
As $T_c$ is low we do not reach very far below the low temperature limit of $T_c$/3; nevertheless, in the accessible temperature range the data are well described by a BCS-like exponential fit (red).
A gap ratio of $\Delta_0/T_c = 1.76$ provides an excellent fit to the data, suggesting a full, isotropic BCS-like superconducting gap. 

Recent theoretical studies \cite{Sasaki-Fu-Ando-PRL-PCS-in-SnInTe,FuBerg-PRL-CBS-parity-theory} show that only three pairing symmetries are possible that do not spontaneously break any lattice symmetry, namely the $A_{1g}$, $A_{1u}$, and $A_{2u}$ representations of $D_{3d}$.
$A_{1g}$ is even parity and fully gapped and corresponds to the $s$-wave state that does not allow topological behavior.
$A_{1u}$ is odd parity and fully gapped; $A_{2u}$ is odd parity and has symmetry-protected point nodes.
Our TDO measurements exclude the $A_{2u}$ parity and point to one of the two fully-gapped states.
If there is unconventional superconductivity in \SnxInxTe, it must be the $A_{1u}$ state, consistent with band structure arguments \cite{Sasaki-Fu-Ando-PRL-PCS-in-SnInTe} that suggest that the pairing symmetry has odd parity.
Recent Knight shift measurements \cite{Maeda-Ando-PRB-SnInTe-NMR-says-not-triplet} on a polycrystalline sample with 4\% In-doping yielded an incomplete suppression of the Knight shift that was nevertheless larger than the expected value for spin-triplet pairing.
These results were interpreted as signature of spin-singlet behavior.
However, since the doping level of this sample is right at the cubic-rhombohedral transition, further studies on higher-doped single-crystals may be needed to obtain a definite answer.
More exotic pairing symmetries would be allowed if evidence of rotational symmetry breaking is seen, as is the case in the doped Bi$_2$Se$_3$ family of superconductors \cite{Matano-NatPhys-CBS, deVisser-SciRep,Asaba-PRX-NBS}.

\begin{figure}
	\includegraphics[width=1\columnwidth]{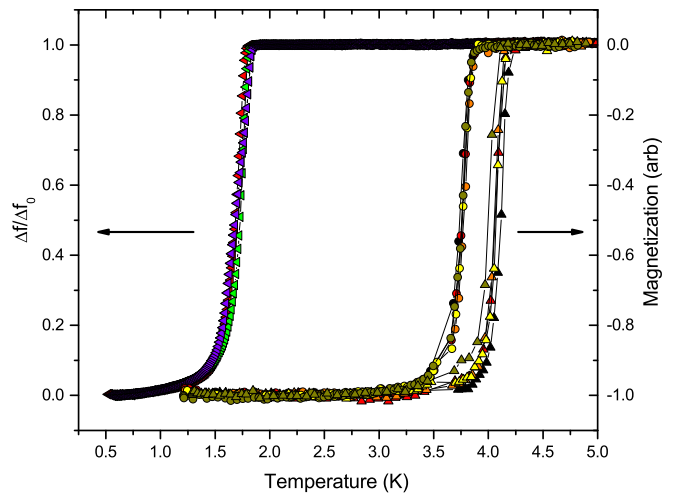}
	\caption{
		Superconducting transitions following repeated irradiations with 5 MeV protons in a crystal of underdoped Sn$_{0.9}$In$_{0.1}$Te ($T_c$ = 1.8 K) and two crystals of near-optimally doped Sn$_{0.55}$In$_{0.45}$Te ($T_c$ = 3.8 K, 4.1 K) as measured by TDO and SQUID magnetometry, respectively.
		With doses up to 2x10$^{17}$ p/cm$^2$, there is essentially no change in the transition temperature, suggesting that adding disorder does not kill competing interactions.
	}
	\label{figIrradiation}
\end{figure}

An open question is the effect of disorder scattering in \SnxInxTe.
TDO and SQUID magnetometry measurements following repeated irradiation with 5 MeV protons up to high total doses of 2x10$^{17}$ p/cm2 on three crystals of \SnxInxTe~with different doping levels are shown in Fig. \ref{figIrradiation}.
With increasing dose, the transition temperature remains essentially constant for all doping levels studied.
If additional scattering were to enhance $T_c$ by suppressing the competing ferroelectric interaction \cite{Novak-Ando-PRB-SnInTe-phase-diagram} in Sn$_{0.9}$In$_{0.1}$Te, an enhancement in $T_c$ should be visible in contrast to our experimental findings on samples with similar doping level as examined in Ref. \onlinecite{Novak-Ando-PRB-SnInTe-phase-diagram}.
A proton dose of 2x10$^{17}$ p/cm$^2$ is substantial and causes clear suppression of $T_c$ in many superconductors.
For instance, in fully-gapped, optimally doped Ba(Fe$_{1-x}$Co$_x$)$_2$As$_2$ proton irradiation suppresses $T_c$ at a rate $\sim$ 5 \% / 10$^{16}$ p/cm$^2$ \cite{Nakajima-PRB2010-Tc-suppression-in-BaFeCoAs-by-protons}, and in fully-gapped, optimally doped Ba$_{1-x}$K$_x$Fe$_2$As$_2$ proton irradiation suppresses $T_c$ at a rate $\sim$ 0.8 \% / 10$^{16}$ p/cm$^2$ \cite{TaenTamegai-Tc-suppression-with-protons-in-BaKFeAs}.
Anderson’s theorem \cite{AndersonTheory} states that $s$-wave superconductivity in isotropic materials remains unaffected by non-magnetic disorder scattering.
Furthermore, recently it has been recognized that due to strong spin-orbit coupling effects, $T_c$ in topological superconductors is surprisingly insensitive to non-magnetic disorder \cite{FuMichaeli-PRL-Robust-odd-parity,Nagai-PRB-CBS-robust,Smylie-PRB-NBS-disorder} regardless of superconducting gap structure.
Thus, the results presented here are consistent with either $A_{1g}$ or $A_{1u}$ symmetry.  

\begin{center}
\textbf{IV. CONCLUSION}
\end{center}

In summary, we have investigated the superconducting properties of the topological crystalline insulator-derived superconductor \SnxInxTe, and have shown it to be a fully-gapped superconductor for x $\geq$ 0.10 with anisotropy increasing with doping.
Magnetic phase diagrams have been extended to $<$ 1K.
One of the two suggested types of odd-parity pairings ($A_{2u}$) cannot describe this material as our results rule out nodal behavior, and the reports of unconventional superconductivity in the material are thus only consistent with the $A_{1u}$ pairing, making \SnxInxTe~ a strong candidate for a topological superconductor.
Proton irradiation does not enhance $T_c$ at any studied doping level, indicating that increasing scattering does not enhance $T_c$ by destroying possible competing ferroelectric interactions or odd parity pairing in the cubic phase.
To fully investigate the interplay of ferroelectricity and superconductivity (and the possibility of competition between odd-parity vs even parity superconductivity), further studies on samples with lower doping will be necessary.

\begin{center}
\textbf{ACKNOWLEDGMENTS}
\end{center}
TDO and magnetization measurements were supported by the U.S. Department of Energy, Office of Science, Basic Energy Sciences, Materials Sciences and Engineering Division.
SANS measurements were supported by the U.S. Department of Energy, Office of Basic Energy Sciences, under Award No. DE-SC0005051.
M.P.S. thanks ND Energy for supporting his research and professional development through the NDEnergy Postdoctoral Fellowship Program.
Work at Brookhaven was supported by the Office of Basic Energy Sciences (BES), Division of Materials Science and Engineering, U.S. Department of Energy (DOE), through Contract No. DE-SC0012704.
R. D. Z. and J. S. were supported by the Center for Emergent Superconductivity, an Energy Frontier Research Center funded by BES.
The work at ORNL was supported by the US Department of Energy, Basic Energy Sciences, Materials Sciences and Engineering Division.
We acknowledge that crystal growth was enabled by individuals such as of M. Susner.

\bibliography{SnInTe-v0.81}

\end{document}